\documentclass[11pt,a4paper]{article}

\usepackage{jheppub}
\usepackage[utf8]{inputenc}
\usepackage{amsmath}
\usepackage{amsthm}
\usepackage{amssymb}
\usepackage{enumitem}   
\usepackage{url}
\usepackage{mathtools}
\usepackage{slashed}
\usepackage{multirow}
\usepackage{xcolor}
\usepackage{slashed}
\usepackage{multirow}
\usepackage[toc,page]{appendix}
\usepackage{hyperref}

\usepackage{url}

\usepackage{amsmath, amsfonts, amssymb}
\usepackage{graphicx}
\usepackage{color}
\usepackage{enumerate}
\usepackage{hyperref}
\usepackage{latexsym}
\usepackage{enumerate}
\usepackage{soul}
\usepackage[normalem]{ulem}
\usepackage{wasysym}
\usepackage{makecell}
\usepackage{slashed}
\usepackage{comment,verbatim}
\usepackage{amsmath}
\usepackage{subfig}

\newcommand{\be}{\begin{equation}}
\newcommand{\ee}{\end{equation}}

\begin{document}



\title{Monodromic transparency of axion domain walls}
\author[]{Simone Blasi}
\affiliation[]{Deutsches Elektronen-Synchrotron DESY, Notkestr.~85, 22607 Hamburg, Germany}
\emailAdd{simone.blasi@desy.de}


\abstract{
We revisit the study of light interacting with QCD axion domain walls from the perspective of the non--linear axion coupling to photons, $g(a) F \tilde F$, which encodes the effects related to the breaking of the axion shift symmetry including the well--known mixing with meson states. As the axion makes an $\mathcal{O}(1)$ excursion of its fundamental period around strings and domain walls, the standard linear coupling to photons is generally insufficient to accurately describe the interaction of light with the defects, and one needs to consider the full structure of $g(a)$. We take this into account in evaluating the friction experienced by axion domain walls moving in a thermal bath of photons, as well as in deriving the birefringent properties of the walls. This clarifies some results in the literature dealing with a special cancellation that takes place for the QCD axion with the electromagnetic and color anomaly as predicted by minimal Grand Unified Theories.}

\preprint{
\begin{flushright}
DESY-24-192
\end{flushright}
}

\maketitle

\section{Introduction}

The interaction of photons with new light pseudoscalar particles is predicted in many scenarios of new physics beyond the Standard Model. The main motivation for such new states comes from the Peccei--Quinn (PQ) solution to the strong CP problem\,\cite{Peccei:1977hh,Weinberg:1977ma,Wilczek:1977pj,Kim:1979if,Shifman:1979if,Zhitnitsky:1980tq,Dine:1981rt}, where the QCD axion couples to photons as a result of the model--dependent chiral anomaly of the PQ current with electromagnetism as well as the model--independent contribution from the strong sector. Together with the basic interaction with axion particle excitations, photons can also interact with the defects that are allowed by the topological properties of the PQ theory, namely axion strings and domain walls\,\cite{Vilenkin:1982ks,Sikivie:1982qv}.

In the following, we will focus on light scattering off QCD axion domain walls. 
This subject has been studied in detail by Huang and Sikivie in Ref.\,\cite{Huang:1985tt}, where it was first recognized that the pion component of the axion walls can play an important role in determining their electromagnetic properties 
(see also\,\cite{Bardeen:1977bd,Sikivie:1984yz,Ganoulis:1986rd,Wilczek:1987mv,PhysRevD.41.1231,Harari:1992ea} for early studies of axion electrodynamics). There it was noticed that the walls become transparent to photons with large enough wavelength, leading to an exponentially suppressed friction from thermal photons at low temperatures, $\Delta P \propto e^{-m_a/T}$, where $m_a$ is the axion mass. 

As the phenomenological implications connected to axion and axion--like particle (ALP) domain walls in the early Universe (including gravitational wave emission\,\cite{Hiramatsu:2012sc,Higaki:2016jjh,Chiang:2020aui,ZambujalFerreira:2021cte,Gelmini:2021yzu,Blasi:2022ayo,Bai:2023cqj,Blasi:2023sej,Kitajima:2023cek,Lu:2023mcz,Ferreira:2024eru,Lee:2024toz}, production of dark matter\,\cite{Gelmini:2022nim,OHare:2021zrq,Gorghetto:2022ikz,Beyer:2022ywc}, baryogenesis\,\cite{Daido:2015gqa,Mariotti:2024eoh}, and primordial black--hole formation\,\cite{Ferrer:2018uiu,Gelmini:2023ngs,Dunsky:2024zdo}) are tied to the dynamics of the network, a precise determination of the domain wall interaction with the thermal plasma is needed for accurate predictions.

In this paper, we will revisit the computation of Ref.\,\cite{Huang:1985tt} to clarify the origin of the axion--wall transparency to photons at low energies, and to determine the actual temperature dependence of the thermal friction in the thin--wall regime, $T \ll m_a$, see also Ref.\,\cite{Hassan:2024bvb} for recent work. This will also explain the discrepancy between Ref.\,\cite{Huang:1985tt} and the preprint version of the same paper, Ref.\,\cite{Huang:prep}, where a different behavior for the thermal friction was suggested, namely $\Delta P \propto T^4$.

As we shall see, the wall transparency originates from a special cancellation between the QCD axion and the pion field that takes place whenever the ratio of the electromagnetic and color anomaly, $E/N$, is taken accordingly to the prediction from minimal Grand Unified Theories (GUTs), namely $E/N = 8/3$
(see e.g.\,\cite{Agrawal:2022lsp}). This particular value of $E/N$ corresponds to the one adopted in Ref.\,\cite{Huang:1985tt}, while the slightly different coupling considered in the preprint\,\cite{Huang:prep} spoils this delicate cancellation, explaining the qualitative discrepancy between the two results. 
However, by evaluating the leading order pressure at low temperatures we will conclude that, even in the case where the cancellation does occur, the photon friction still follows a power law, namely $\Delta P \sim \alpha^2 (T^2/m_a)^4$.

Our calculations will be conveniently carried out by referring to the non--linear, monodromic, axion coupling to photons, $g(a) F \tilde F$, as introduced in Ref.\,\cite{Agrawal:2023sbp}, (see also\,\cite{Agrawal:2017cmd,Fraser:2019ojt} for previous related work) which is well suited for this kind of problem. 
In fact, the coupling $g(a)$ describes the low--energy dynamics of axions and photons including the effects related to the breaking of the axion shift symmetry. In particular, the non--linearities of $g(a)$ correlate with the axion potential and can thus capture the mixing with the QCD mesons. 
Such non--linearities become particularly important when considering axion defects such as strings and walls, where the axion makes an $\mathcal{O}(1)$ excursion of its fundamental period.
By comparing the predictions obtained with the full coupling $g(a)$ against the ones from its linearized version, which is customarily parameterized by $\frac{1}{4} g_{a \gamma \gamma} a F \tilde F$, we will show how the latter does not in general provide an accurate description of the defect electrodynamics.

In this regard, we will also revisit the birefringent properties of axion strings and walls, which are known to be sensitive to meson mixing\,\cite{Agrawal:2019lkr}, by comparing the results obtained with $g(a)$ against the standard linear coupling, with the latter being most commonly used in this context, see e.g.\,\cite{Takahashi:2020tqv,Jain_2021,Kitajima_2022,Gonzalez:2022mcx,Jain_2022,Ferreira:2023jbu,Hagimoto:2024sgw,Lee:2024xjb}
\footnote{Non--linearities may be relevant in other scenarios as well where the axion field excursion between the photon emission and absorption is large, as for instance for light traveling in a dense axion environment around black holes\,\cite{Gan:2023swl}.}.

Finally, in Sec.\,\ref{sec:variations} we will discuss two variations of the minimal QCD axion setup to distinguish the generic features of axion--wall electrodynamics from accidental cancellations.

\section{Light scattering off axion domain walls}
\label{sec:setup}

In this section we derive the relevant interactions between photons and axions/pions, which will be used to determine the dynamics of light scattering off axion domain walls.

\medskip
The relevant part of the Lagrangian is given by
\be
\label{eq:UVL}
\mathcal{L} = - \frac{1}{4} G^{\,\text{a} \, \mu \nu}  G_{\mu \nu}^{\,\text{a}} - \frac{1}{4} F^{\mu \nu} F_{\mu \nu} + \frac{1}{2} F_a^2 (\partial_\mu a)^2 + a \, N \frac{\alpha_s}{4 \pi} G^{\,\text{a} \, \mu \nu}  \tilde G_{\mu \nu}^{\,\text{a}} + a \, E \frac{\alpha}{4 \pi} F^{\mu \nu} \tilde F_{\mu \nu},
\ee 
where $N$ and $E$ indicate the color and the electromagnetic anomaly of the PQ current, respectively, and we have introduced the axion as a dimensionless field with fundamental period $F_a$. Derivative couplings of the axion to SM fermions may be present alongside \eqref{eq:UVL}.
The color anomaly is related to the domain wall (DW) number via $N_\text{DW} = 2 N$. The dual field strength is defined by
$\tilde G_{\mu \nu}^\text{\,\text{a}} = \frac{1}{2} \epsilon_{\mu \nu \rho \sigma} G^{\,\text{a} \, \rho \sigma}$, and similarly for $\tilde F_{\mu \nu}$.

We will follow the approach by which the linear coupling of the axion to $G \tilde G$ is rotated away by a chiral transformation of the light quark fields, see e.g.\,\cite{GrillidiCortona:2015jxo}, leading to: 
\be
\mathcal{L} = - \frac{1}{4} G^{\,\text{a} \, \mu \nu}  G_{\mu \nu}^{\,\text{a}} - \frac{1}{4} F^{\mu \nu} F_{\mu \nu} + \frac{1}{2} F_a^2 (\partial_\mu a)^2 + a \frac{\alpha}{4 \pi} \left[ E - 6 N \,\text{Tr}\,(Q_a Q^2) \right] F^{\mu \nu} \tilde F_{\mu \nu},
\ee 
where $Q$ encodes the electromagnetic quark charges,
\be
Q = \begin{pmatrix} \frac{2}{3} & 0\\
0 & - \frac{1}{3}
\end{pmatrix}.
\ee
Due to the rotation performed on the quark fields, the quark mass matrix now depends on the axion as:
\be
M_a = e^{i a N Q_a} \begin{pmatrix} m_u & 0\\
0 & m_d
\end{pmatrix}
e^{i a N Q_a}, \quad \,\text{Tr}\,Q_a = 1.
\ee
\medskip

At low energies, the interaction of the axion with photons and QCD bound states can be described within chiral perturbation theory ($\chi$PT). At leading order in $\chi$PT, one has:
\be
\label{eq:LO}
\mathcal{L}_\text{LO} =  \frac{f_\pi^2}{4} \left[ \langle D_\mu U^\dagger D^\mu U \rangle +
2 B_0 \langle U M_a^\dagger + M_a U^\dagger \rangle \right] + a \frac{\alpha}{4 \pi} \left[ E - 6 N \,\text{Tr}\,(Q_a Q^2) \right] F^{\mu \nu} \tilde F_{\mu \nu} + \mathcal{L}_\text{WZW},
\ee
where $\langle \dots\rangle$ indicates a trace in flavor space, and we have introduced a dimensionless two--flavor pion matrix as
\be
U = e^{ i \Pi }, \quad \Pi = \begin{pmatrix} \pi^0 & \sqrt{2} \pi^+ \\
\sqrt{2} \pi^- & - \pi^0
\end{pmatrix},
\ee
as well as the Wess--Zumino--Witten (WZW) term\,\cite{Wess:1971yu,Witten:1983tw} to account for the coupling of the pions to photons. 

As it is well known, the axion and the neutral pion interact via the potential term in \eqref{eq:LO} such that axion domain wall solutions will involve a non--trivial $\pi^0$ profile as well. Let us then focus on the $a-\pi^0$ system. From the WZW term one obtains, see e.g.\,\cite{Lu:2020rhp}, 
\be
\mathcal{L}_{\pi^0 \gamma \gamma} = i \frac{\alpha}{4 \pi} \epsilon^{\mu \nu \rho \sigma}
\partial_\nu A_\rho A_\sigma
\langle 2 Q^2 ( U \partial_\mu U^\dagger - U^\dagger \partial_\mu U) - Q U^\dagger Q \partial_\mu U^\dagger + Q U Q \partial_\mu U^\dagger \rangle,
\ee
which evaluates to
\be
\mathcal{L}_{\pi^0 \gamma \gamma} = \frac{\alpha}{4\pi} \pi^0 F^{\mu \nu} \tilde F_{\mu \nu}.
\ee
The dynamics of axions, neutral pions, and photons is then described by
\be
\label{eq:apigamma}
\mathcal{L}_{a \pi} = \frac{1}{2} F_a^2 (\partial_\mu a)^2 + \frac{1}{2} f_\pi^2 (\partial_\mu \pi^0)^2 - V(a,\pi^0)_{Q_a}  +  \frac{\alpha}{4\pi} \left[ \pi^0  + a \left( E - 6 N \,\text{Tr}\,(Q_a Q^2) \right) \right] F^{\mu \nu} \tilde F_{\mu \nu},
\ee
where $V(a, \pi^0)_{Q_a}$ is the potential term coming from \eqref{eq:LO}, and we have stressed the dependence on the specific choice for the matrix $Q_a$.

The coupling of the physical axion to photons can be obtained directly from \eqref{eq:apigamma} by appropriately choosing $\bar Q_a = M_q^{-1}/ \langle M_q^{-1} \rangle$, such that no mixing between $a$ and $\pi^0$ ever arises from the potential $V(a, \pi^0)_{\bar Q_a}$ around the vacuum at $(a, \pi^0)=(0, 0)$. For generic choices of $Q_a$ one needs to rotate the fields to their mass eigenstates, and include the contribution from the WZW term involving the $\pi^0$ as well. The final result is nevertheless independent of $Q_a$, and one obtains the well known expression at leading--order in $\chi$PT:
\be
\label{eq:linearcoupling}
\mathcal{L}_{a \pi} \supset \frac{1}{4} g_{a \gamma \gamma} a_\text{phys}  F^{\mu \nu} \tilde F_{\mu \nu}, \quad 
g_{a \gamma \gamma} = \frac{\alpha}{2 \pi f_a} \left( \frac{E}{N} - \frac{2}{3} \frac{4 m_d + m_u}{m_d+m_u} \right),
\ee
where $f_a \equiv F_a/2 N = F_a/N_\text{DW}$, and we have indicated by $a_{\rm phys}$ the physical axion state.

Eq.\,\eqref{eq:apigamma} can be used to derive the Maxwell equations in the background of an axion/pion field. Let us then consider the case of a planar domain wall solution characterized by the profiles $a(z)$ and $\pi^0(z)$. One has: 
\be
\label{eq:gammaDW}
\partial_\mu F^{\mu \nu} = \frac{\alpha}{\pi} \partial_\mu \beta(z)  \tilde F^{\mu \nu},
\ee
where we have introduced
\be
\label{eq:betadef}
\beta(z) \equiv  \pi^0(z) + \left[ E - 6 N \,\text{Tr}(Q_a Q^2) \right] a(z).
\ee
In order to study the interaction of photons with the domain walls via \eqref{eq:gammaDW}, one needs some information on the profiles $a(z)$ and $\pi^0(z)$.

At this point let us make contact with the original work of Huang and Sikivie\,\cite{Huang:1985tt}, who considered a model with $N_\text{DW}=2$ ($N=1$) and performed the computation in the same two--flavor scheme with $Q_a = \mathbb{I}/2$. In this case the potential reads:
\be
\label{eq:Videntity}
V(a, \pi^0)_{\mathbb{I}/2} = - m_\pi^2 f_\pi^2 \left[ \frac{z}{1+z} \,\text{cos}(a - \pi^0) + \frac{1}{1+z} \,\text{cos}(a + \pi^0) \right], \quad z \equiv m_u/m_d,
\ee
and
\be
\label{eq:betaidentity}
\beta(z) = \pi^0(z) + \left( E - \frac{5}{3} N\right) a(z).
\ee
This potential allows for two degenerate, physically distinct, vacua located at
\be
\, \text{L}\, =(a, \pi^0) = (0,0), \quad \, \text{R}\, =(a, \pi^0) = (\pi,-\pi),
\ee
with domain wall solutions interpolating between L at $z=-\infty$ and R at $z= + \infty$ (and vice versa). 
By looking at the value of $\beta(z)$ in \eqref{eq:betadef}, one finds
\be
\label{eq:betaLR}
\beta(-\infty) \equiv \beta_\text{L} = 0, \quad  \beta(\infty) \equiv \beta_\text{R} = \left( \frac{E}{N} - \frac{8}{3} \right) \pi.
\ee
As we can see, whenever the anomaly coefficients are such that $E/N = 8/3$ as predicted by minimal GUTs, the function $\beta^\prime(z)$ that couples the photons to the axion domain walls in \eqref{eq:gammaDW} actually averages out across the wall, as $\beta_\text{L} = \beta_\text{R}$. A photon with a wavelength that is much larger than the wall width will not (at leading order) notice the presence of the wall in this case.

Ref.\,\cite{Huang:1985tt} did consider this particular coupling with $E/N = 8/3$, and noticed that the reflection probability for a photon scattering off the wall could be sizable only for frequencies $\omega \sim \delta_w^{-1} \sim m_a$, where $\delta_w$ indicates the wall width. On the other hand, a slightly different coupling was considered in the preprint version of the same paper\,\,\cite{Huang:prep}\,\footnote{The axion coupling to photons in\,\cite{Huang:prep} reads $\frac{\alpha}{3\pi} (2N) a F \tilde F$, which would correspond only to the pure $E$ term in Eq.\,\eqref{eq:LO} when adopting the minimal GUT relation $E/N = 8/3$.} which actually prevents the cancellation in\,\eqref{eq:betaLR} to occur, explaining the qualitatively different results.

The suppressed photon--wall interaction for $E/N = 8/3$ hinges on the non--trivial pion profile. This effect is not visible if one considers only the linear axion--photon coupling for the mass eigenstate, $g_{a \gamma \gamma}$ in Eq.\,\eqref{eq:linearcoupling}, which in fact shows no particular cancellation for this value of $E/N$. The reason is that the linear axion coupling only works in the vicinity of the vacuum at $(a,\pi^0) = (0,0)$, whereas the domain wall solution involves an $\mathcal{O}(1)$ excursion of the fundamental period for both the axion and the pion field.

\medskip
In the next section, we will show how the interaction of light with axion domain walls can be captured in the effective field theory (EFT) where the pion is integrated out by referring to the non--linear monodromic axion coupling to photons as introduced in Ref.\,\cite{Agrawal:2023sbp}. This will streamline the calculation of the relevant scatterings by reducing it to a one--field problem, as well as showcase the importance of the non--linearities of the axion coupling when studying defects such as axion strings and walls.

\subsection{Photon--wall scattering in the EFT}
\label{sec:EFT}

An important property of the axion/pion domain wall solution allowed by the potential in \eqref{eq:Videntity} is that both fields vary on scales of order $\sim m_a^{-1}$ for $F_a \gg f_\pi$\,\cite{Huang:1985tt,Forbes:2000et}. It is then reasonable to integrate out the pion 
field at the tree level in the limit $m_\pi^2 \gg m_a^2$ by solving
\be
\label{eq:integrateout}
\frac{ \partial}{\partial \pi^0}  V(a,\pi^0)_{Q_a}= 0.
\ee
Let us consider $z \equiv m_u/m_d < 1$ as it is the case for QCD. By choosing $Q_a^\prime = \,\text{diag}\,(1,0)$ we can solve \eqref{eq:integrateout} as
\be
\label{eq:pi0sol}
\pi^0 = \,\text{tan}^{-1} \frac{ \,\text{sin}\,( 2 N a)}{z^{-1} + \,\text{cos}\,( 2 N a)}.
\ee
Notice that because of this choice, the potential $V(a,\pi^0)_{Q_a^\prime}$ is such that the pion field will be vanishing on both sides of the axion wall.

The low energy effective theory is then given by\,\cite{Agrawal:2023sbp}:
\be
\mathcal{L}_{a \gamma} = - \frac{1}{4} F^{\mu \nu} F_{\mu \nu} + \frac{1}{2} F_a^2 (\partial_\mu a)^2 - V(a) + \frac{\alpha}{4\pi} g(a) F^{\mu \nu} \tilde F_{\mu \nu},
\ee
where
\be
\label{eq:myga}
g(a) = \,\text{tan}^{-1} \frac{ \,\text{sin}\,( 2 N a)}{z^{-1} + \,\text{cos}\,( 2 N a)} + \left( E -  \frac{8}{3} N \right) a,
\ee
and
\be
\label{eq:Vofa}
V(a) = - m_\pi^2 f_\pi^2 \sqrt{ 1 - \frac{4 z}{(1+z)^2} \,\text{sin}^2(N a )}.
\ee
As we can see, the non--linear coupling $g(a)$ is simply given in this case by \eqref{eq:betadef} with $Q_a = \,\text{diag}\,(1,0)$ when replacing $\pi^0(z)$ with the profile in \eqref{eq:pi0sol}. 

\begin{figure}
\centering
\hspace{-0.4cm}
 \includegraphics[scale=0.25]{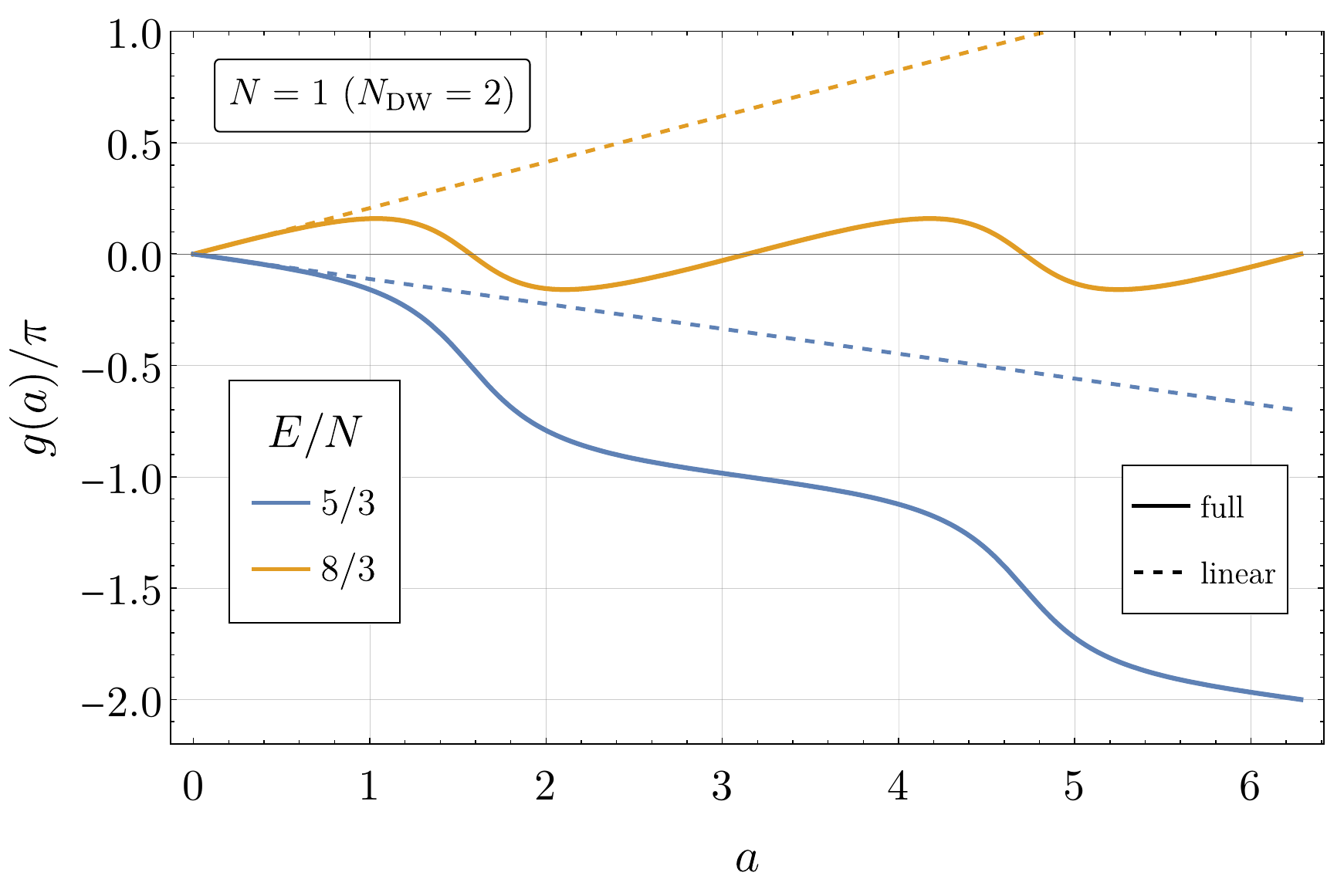}
 \caption{The function $g(a)$ in Eq.\,\eqref{eq:myga} for typical values of $E/N$, and its linearized version around $a=0$.}
\label{fig:gofa} 
\end{figure}

The function $g(a)$ is shown in Fig.\,\ref{fig:gofa} for two characteristic values of $E/N$ including $8/3$, together with its linearized version, $g^\prime(0)a$, corresponding to the standard linear coupling $g_{a \gamma \gamma} a_{\rm phys}$ in \eqref{eq:linearcoupling}, for comparison.

The potential $V(a)$ has degenerate minima at $a_k = \pi k/N$ for $k=0,\dots, 2 N -1$, and domain wall solutions can interpolate between $a_k$ and $a_{k+1}$. On the two sides of the wall, the coupling $g(a)$ is such that
\be
\label{eq:deltagDW}
\Delta[g(a)]_{\text{DW}} \equiv g(a_{k+1}) - g(a_{k}) = \left(  \frac{E}{N} - \frac{8}{3} \right) \pi,
\ee
which is a generalization of \eqref{eq:betaLR} for arbitrary $N$. 
The monodromic charge $n$, given as 
\be
\label{eq:monocharge}
g(a + 2\pi) - g(a) = \left(E - \frac{8}{3} N\right) 2 \pi \equiv 2 \pi n,
\ee
is in agreement with\,\cite{Agrawal:2023sbp} as we use a different but equivalent form of $g(a)$ in Eq.\,\eqref{eq:myga}.

The equations of motion for light scattering off the wall are now given by the following equivalent form of \eqref{eq:gammaDW}:
 \be
\partial_\mu F^{\mu \nu} = \frac{\alpha}{\pi} \, \partial_\mu \left[ g(a) \right] \, \tilde F^{\mu \nu}.
\ee 
Let us then consider a photon traveling along the $z$ direction with frequency $\omega$. In the Lorentz gauge $\partial_\mu A^\mu = 0$ this leads to the familiar equation for the $\pm$ photon helicity scattering off the wall located around $z=0$ (see e.g.\,\cite{Ganoulis:1986rd,Favitta:2023hlx} for a derivation):
\be
\label{eq:eoms}
(\partial_z^2 + \omega^2) A_{\pm}(z) = \mp \frac{\alpha}{\pi} \omega \,
\frac{ \text{d}}{\text{d}z} [g(a)] \,
A_{\pm}(z),
\ee
where we have introduced $A_\pm = A_x \pm i A_y$.

Let us now evaluate the reflection probability $\mathcal{R}_\pm$ in the Born approximation. At the leading order in $\alpha$ one has: 
\be
\label{eq:Born}
A_\pm(z) \sim e^{i \omega z}  \pm i \frac{\alpha}{2\pi} \int_{-\infty}^{+\infty} \text{d}z^\prime e^{i \omega | z - z^\prime|} \frac{ \text{d}}{\text{d}z^\prime} [g(a)] \, e^{i \omega z^\prime} + \mathcal{O}(\alpha^2),
\ee
leading to
\be
\label{eq:Rpm}
\mathcal{R}_\pm =  \frac{\alpha^2}{4 \pi^2} \bigg| \int_{-\infty}^{+\infty} \text{d} z^\prime \, \frac{\text{d}}{\text{d}z^\prime}[g(a)] e^{2 i \omega z^\prime}\bigg|^2.
\ee
Notice that the reflection probability cannot distinguish the photon helicity at the leading order in $\alpha$. In the kinematic limit mentioned at the end of Sec.\,\ref{sec:setup}, where $\omega \ll \delta_w^{-1} \sim m_a$, one simply has:
\be
\label{eq:Rgenericlow}
\mathcal{R}_\pm (\omega \ll m_a) \simeq  \frac{\alpha^2}{4 \pi^2} \Delta[g(a)]^2_\text{DW} = \frac{\alpha^2}{4} \left(  \frac{E}{N} - \frac{8}{3} \right)^2.
\ee 

This expression agrees with the observation in Ref.\,\cite{Huang:1985tt} that the reflection probability is actually suppressed when $\omega \ll m_a$ for the minimal GUT prediction, $E/N = 8/3$. For this special value, one needs to further expand \eqref{eq:Rpm} for $\omega z^\prime \ll 1$ to find the first non--zero contribution:
\be
\label{eq:R83low}
\mathcal{R}_\pm^{8/3} (\omega \ll m_a) \simeq c \cdot \alpha^2 \left(\omega/m_a\right)^4, \quad c \simeq 2.16..
\ee
where $c$ is a numerical $\mathcal{O}(1)$ coefficient. For this explicit calculation we have taken for simplicity the domain wall profile as if the potential was a perfect cosine rather than \eqref{eq:Vofa},
namely $a(z) = (2/N)\,\text{tan}^{-1}\left(e^{m_a z}\right)$, as this will only change the precise numerical value of $c$.
As anticipated in the previous section, reflection is indeed strongly suppressed at low energies, as the first non--zero contribution scales like $ \mathcal{R} \sim \alpha^2 (\omega/m_a)^4$\,\footnote{A less suppressed behavior with $\omega/m_a$ may arise at higher orders in $\alpha$, which could be dominant at very low energies.}.

In the opposite limit, $\omega \gg m_a$, reflection is exponentially suppressed as the width of the wall becomes much larger than the photon wavelength. We find the reflection probability to be well approximated by:
\be
\mathcal{R}_\pm (\omega \gg m_a) \simeq  \frac{\alpha^2}{2\pi} e^{- \delta_w \omega}, \quad \delta_w = \sqrt{2} \,m_a^{-1}.
\ee 
Notice that this result does not depend on $E/N$ at the leading order. The reason is that the (non--linear) pion component of $g(a)$ in \eqref{eq:myga} is the one keeping the photon interacting with the axion wall for the highest momenta.

At this point it is apparent that the linear axion coupling $g^\prime(0) a F \tilde F$ would lead to different results both at the quantitative and qualitative level when studying photons scattering off the wall, as it would only account for the axion--pion mixing around the minimum of the potential while neglecting the pion profile altogether\,\footnote{Alternatively, one can of course still work within the two--field setup, and take into account the pion field as done originally in Ref.\,\cite{Huang:1985tt} and discussed around Eq.\,\eqref{eq:betaidentity}.}. In particular, the linear coupling can not account for the cancellation at $E/N = 8/3$, while for generic values of $E/N$ the probability will still differ quantitatively,
\be
\mathcal{R}_\pm^\text{linear} (\omega \ll m_a) \simeq \frac{\alpha^2}{4} \left( \frac{E}{N} - \frac{2}{3} \frac{4+z}{1+z}\right)^2,
\ee 
which agrees with \eqref{eq:Rgenericlow} only for $z \equiv m_u/m_d \ll 1$, as the non linearities become less important in this limit. Quantitatively different results are also found in the limit $\omega \gg m_a$, as we shall see below.

\begin{figure}
\centering
\hspace{-0.4cm}
 \includegraphics[scale=0.25]{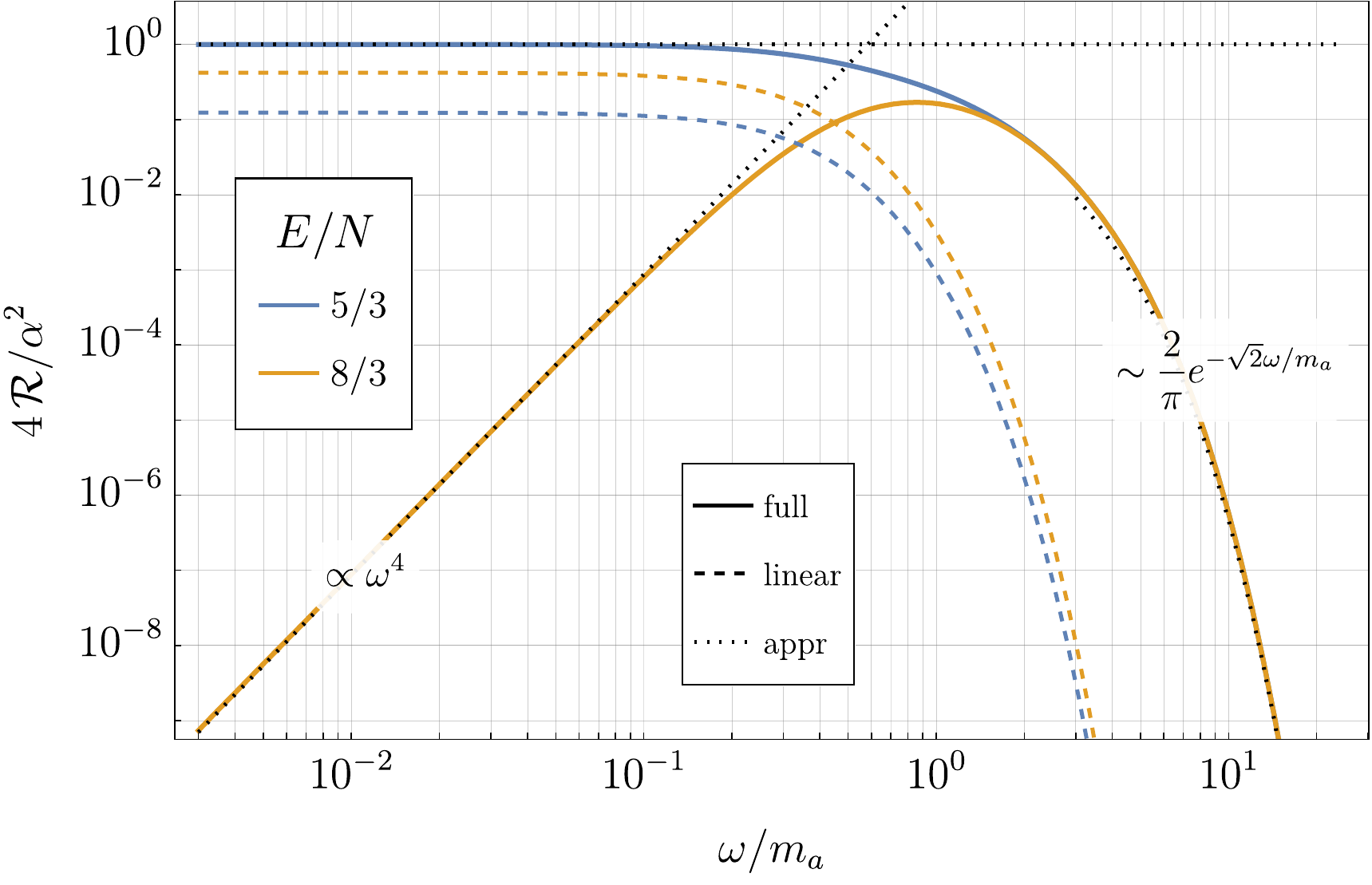}
  \includegraphics[scale=0.26]{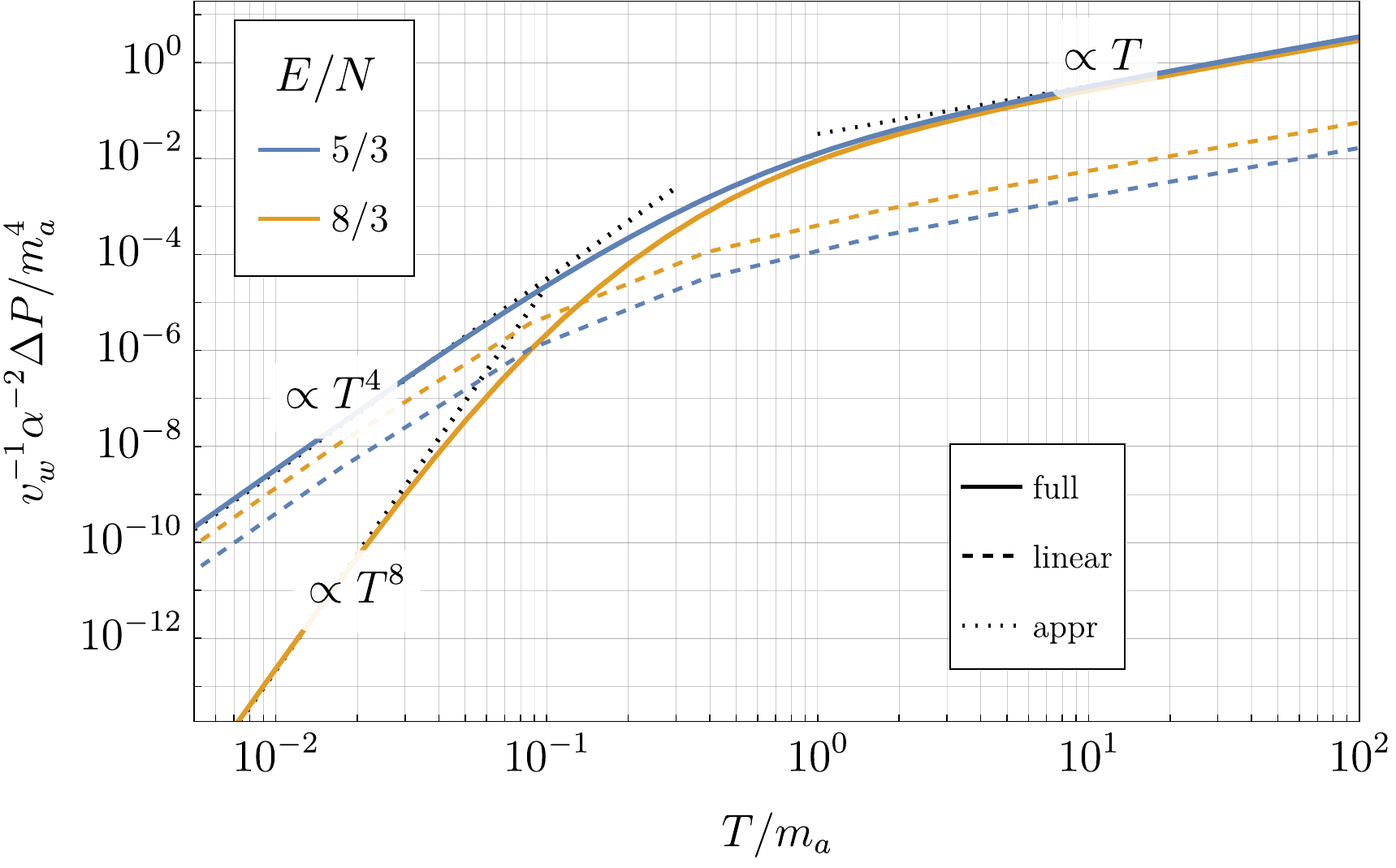}
 \caption{\textbf{Left:} Reflection probability $\mathcal{R}$ normalized to $\alpha^2/4$ for $E/N = 5/3$ (blue) and $E/N = 8/3$ (orange) as a function of the photon frequency. Solid (dashed) lines are obtained by solving numerically the scattering problem with the full coupling $g(a)$ (linearized coupling $g^\prime(0) a$). Dotted lines indicate the analytical approximations derived in Sec.\,\ref{sec:EFT}.  
\textbf{Right:} Thermal pressure on the axion wall as a function of the temperature. The conventions for the plot are the same as in the left panel. The special case with $E/N = 8/3$ implies a pressure $\Delta P \propto T^8$ at low temperatures. For this plot we have taken $v_w = 0.1$.}
\label{fig:fig} 
\end{figure}

The reflection probability as evaluated by solving numerically the equations of motion for the vector potential \eqref{eq:eoms} is shown in the left panel of Fig.\,\ref{fig:fig} as a function of the photon frequency for $E/N = 8/3$ and $E/N = 5/3$. Solid and dashed lines indicate the numerical result with the full $g(a)$ and the linearized coupling $g^\prime(0) a$, respectively, while dotted lines show the analytical approximations derived in this section. The case of $E/N = 8/3$ shows indeed a suppressed reflection probability at low energies and the $\mathcal{R} \sim \alpha^2 (\omega/m_a)^4$ behavior is well reproduced numerically, while for $E/N = 5/3$ one has a constant $\mathcal{R} \sim \alpha^2$. At high energies, the reflection probability turns out to be largely independent of the precise value of $E/N$. As we can see, the linearized couplings provide quantitatively different results in all kinematic regimes, but they qualitatively agree with the case $E/N = 5/3$ where no special cancellation takes place. 

\subsection{Thermal pressure}
\label{sec:thermal}

With the reflection probability at hand, we can now evaluate the pressure on the axion domain wall due to photon scattering by assuming a thermal distribution. Notice that the picture of a single photon interacting with the wall in isolation is justified only if the wall thickness is much smaller than any other thermal scale in the problem, such as the mean free path, as studied in detail in Ref.\,\cite{Hassan:2024bvb}. Strictly speaking the following analysis is therefore applicable only for $T \lesssim m_a$.

The net pressure on a domain wall moving at speed $v_w$ in the photon bath can be written as\,\cite{Arnold:1993wc,Blasi:2022ayo}:
\begin{equation}
\label{eq:pressurea}
    \Delta P = g \frac{2}{(2\pi)^2}
    \int_0^\infty \text{d}\omega \, \mathcal{R}(\omega) 
    \frac{\omega^2}{\beta \gamma a}
    \left[ \,
    2 \beta \, \gamma \, \omega \, v_w - \text{log}\left(\frac{f(-v_w)}{f(v_w)}\right)\right]\bigg|_{E=\omega},
\end{equation}
where $g=2$, $a=-1$ for Bose--Einstein statistics, $\beta=1/T$, and $f(v)$ is the photon thermal momentum distribution in the wall frame. 

By following the approximation employed in Ref.\,\cite{Huang:1985tt} that reflection is only effective for $\omega \sim m_a$, namely $\mathcal{R} = \alpha^2 \omega \, \delta(\omega - m_a)$, one would obtain:
\be
\label{eq:HSlow}
\Delta P^{8/3}_{\rm HS} (T \ll m_a) \simeq \alpha^2 \frac{m_a^3 T}{\gamma \pi^2} e^{-(1-v) \gamma m_a/T} \simeq \frac{\alpha^2}{\pi^2} \, m_a^3 \, T \, e^{-m_a/T},
\ee
where we have additionally taken $\gamma \simeq 1$. 

This result however changes significantly when relaxing the approximation that reflection is only active for $\omega \sim m_a$. For $T \ll m_a$ the pressure is dominated by momenta for which the reflection probability is actually the one given in Eq.\,\eqref{eq:R83low}. This leads to a power law behavior as a function of $T$ rather than the exponential suppression in \eqref{eq:HSlow}, that we evaluate to be 
\be
\Delta P^{\,8/3}(\,T \ll m_a) \simeq  v_w \cdot c \cdot \alpha^2 \frac{2}{\pi^2}  \, \Gamma(8) \,\left( \frac{T}{m_a} \right)^4 T^4 ,
\ee
where $\Gamma(8) = 7!$, and we have again taken the limit of small velocity with $\gamma \simeq 1$. As for the opposite kinematics, $T \gg m_a$, the choice of $E/N = 8/3$ does not have a strong impact, and the pressure will be the generic one as given below in \eqref{eq:DPhigh}.

Let us now turn to generic values of $E/N$. The low--energy reflection probability is the one in\,\eqref{eq:Rgenericlow}, leading to:
\be
\label{eq:DPgeneric}
\Delta P^{\,E/N}(\,T \ll m_a) \simeq v_w \frac{\alpha^2}{2 \pi^2} \left( \frac{E}{N} - \frac{8}{3} \right)^2 \Gamma(4) \, T^4, 
\ee
while at high temperatures one has:
\be
\label{eq:DPhigh}
\Delta P^{\,E/N}(\,T \gg m_a) \simeq v_w \frac{\alpha^2}{\pi^3} m_a^3 \,T.
\ee
As we are now away from the special value $E/N = 8/3$, these expressions qualitatively agree with the preprint in Ref.\,\cite{Huang:prep} as well as with Ref.\,\cite{Hassan:2024bvb} where the linear axion coupling is considered. 

A comparison of the thermal pressure evaluated numerically with the full $g(a)$ and with the linearized coupling is shown in the right panel Fig.\,\ref{fig:fig} by the solid and dashed lines, respectively, for different values of $E/N$. Dotted lines indicate the analytical approximations derived in this section. As we can see, the special case of $E/N = 8/3$ leads to a thermal pressure which is not exponentially suppressed but rather $\propto T^8$ at low temperatures. At high temperatures, the pressure is not particularly sensitive to the value of $E/N$ when evaluated with $g(a)$, similarly to what happens for $\mathcal{R}$ in this regime. The results obtained with the linearized coupling turn out to differ from the full results at all temperatures.

\subsection{Birefringence}
\label{sec:biref}

Let us discuss here the effect of the non--linear coupling $g(a)$ for birefringence, see also\,\cite{Agrawal:2019lkr}. Linearly polarized light propagating in the background of a varying axion field will experience a rotation of the polarization angle that is related to the variation of the axion field along the path (but not on the specific realization of the path). Assuming that the axion field varies slowly over distances of the order of the photon wavelength, the rotation of the polarization angle can be computed in the WKB approximation, leading to\,\cite{PhysRevD.41.1231,Harari:1992ea}:
\be
\label{eq:biref}
\Delta \Phi = \frac{1}{2} g_{a \gamma \gamma} \, \Delta a_\text{phys},
\ee  
where we have considered the standard linear parameterization \eqref{eq:linearcoupling}. Similar relations hold for light passing in the vicinity of topological defects such as axion strings\,\,\cite{Harari:1992ea,Agrawal:2019lkr} where the axion winds around its fundamental period, and $\Delta a_{\rm phys} \sim F_a$.

When the axion mixes with the pion, one needs to include an additional contribution that takes into account the pion field excursion along the path\,\cite{Agrawal:2019lkr}:
\be
\Delta \Phi = \frac{\alpha}{2\pi} \left[ \left( E - \frac{5}{3} N \right)\Delta a + \Delta \pi \right],
\ee
where we have used $Q_a = \mathbb{I}/2$, and assumed that the WKB approximation is applicable to the pion field as well.

Performing a straightforward generalization of \eqref{eq:biref}, the polarization rotation can be evaluated within the same approximation by referring to the non--linear coupling $g(a)$. One simply has:
\be
\label{eq:DphiWKB}
\Delta \Phi = \frac{\alpha}{2\pi} \Delta[ g(a)].
\ee
The advantage is that the non--linear coupling automatically takes into account the contribution from meson mixing. In particular, it is straightforward to show that 
\be
\label{eq:deltaPhistring}
\Delta \Phi_\text{str} =  \pm \frac{\alpha}{2\pi} [ g(a+2\pi) - g(a)] = \pm \, n \cdot \alpha,
\ee
where $n$ is the monodromic charge defined in \eqref{eq:monocharge}, which in fact vanishes for $E/N = 8/3$ as noted in Ref.\,\cite{Agrawal:2023sbp}. A photon looping around such a string would then have its linear polarization unchanged. 

Let us now consider a photon traversing an axion domain wall. In the limit where the photon frequency satisfies $\omega \gg m_a$, the WKB approximation applies leading to
\be
\label{eq:phidw}
\Delta \Phi_{\text{DW}}(\omega \gg m_a) = \frac{\alpha}{2\pi} \Delta[g(a)]_{\text{DW}} = 
\pm \, \frac{\alpha}{2} \left( \frac{E}{N} - \frac{8}{3} \right),
\ee
where we have used \eqref{eq:DphiWKB} and \eqref{eq:deltagDW}. Notice that as a photon looping around the string will traverse exactly $2N$ of these walls, Eq.\,\eqref{eq:deltaPhistring} can be seen as a particular case of \eqref{eq:phidw} with $\Delta \Phi_\text{str} = 2 N \cdot \Delta \Phi_{\text{DW}}$.
\medskip

Let us now make one further step and relax the assumption of a slowly varying axion field compared to the photon wavelength. This is in fact the case of a photon interacting with an axion domain wall in the opposite regime of $\omega \ll m_a$. The rotation of the polarization can then be related to the transmission coefficient for photons with opposite helicity, see e.g.\,\cite{Ganoulis:1986rd}:
\be
\label{eq:rotationtrans}
\Delta \Phi_\text{DW} = \frac{1}{2} \,\text{arg}(T_+/T_-).
\ee
This picture is consistent for $\alpha \ll 1$, as we know from Sec.\,\ref{sec:EFT} that the transmission probability is the same for both helicities at leading order, $\mathcal{T}_\pm \equiv |T_\pm|^2 = 1 - \mathcal{O}(\alpha^2)$, while \eqref{eq:rotationtrans} will be non--zero already at $\mathcal{O}(\alpha)$. 

The imaginary part of $T_+/T_-$ can be evaluated at the leading order in the Born approximation by taking $z \rightarrow \infty$ in \eqref{eq:Born}:
\be
A_\pm(z \rightarrow \infty) \sim e^{i \omega z}  \left[ 1 \pm i \frac{\alpha}{2\pi} \int_{-\infty}^{+\infty} \text{d}z^\prime \frac{ \text{d}}{\text{d}z^\prime} g(a) \right] + \mathcal{O}(\alpha^2),
\ee
leading to
\be
\label{eq:dPhiDW}
\Delta \Phi_\text{DW} = \frac{\alpha}{2\pi} \Delta[g(a)]_\text{DW} = \pm \frac{\alpha}{2} \left( \frac{E}{N} - \frac{8}{3} \right).
\ee
This coincides with \eqref{eq:phidw}, which was obtained within the approximation of a slowly--varying axion field, showing that for the case of light propagating through axion domain walls this assumption can be relaxed.
The rotation angle is then in general $\Delta \Phi \sim \mathcal{A} \, \alpha$ with $\mathcal{A}$ a rational number, while for the specific value $E/N = 8/3$ this vanishes independently of the hierarchy between $\omega$ and the inverse wall width. 

We can compare \eqref{eq:dPhiDW} with the result obtained with the linearized coupling,
\be
\label{eq:dPhiDWlinear}
\Delta \Phi_\text{DW}^\text{linear} = \pm \frac{\alpha}{2} \left( \frac{E}{N} - \frac{2}{3} \frac{4+z}{1+z} \right)
\ee
which gives the same result only in the limit of small $z$. As we can see, the use of the non--linear coupling is important in determining the correct birefringence when considering $O(1)$ excursions for the axion field. 

Let us finally notice that the rotation of the polarization angle \eqref{eq:dPhiDW} does distinguish between different values of $E/N$ even for $\omega \gg m_a$, whereas the reflection probability as well as the thermal pressure (the latter for $T \gg m_a$) are actually the same in this limit, as shown in Fig.\,\ref{fig:fig}.

\section{Variations}
\label{sec:variations}

In this section we discuss two possible variations of the setup presented in Sec.\,\ref{sec:setup} to identify the generic features of the axion--pion domain wall system, as opposed to the properties that are instead sensitive to the specific model. In Sec.\,\ref{sec:mumoremd} we will consider a scenario where the up quark is heavier than the down quark, while in Sec.\,\ref{sec:heavy} we will discuss the case of a heavy QCD axion for which the dominant contribution to the mass comes from an additional confining sector aligned with QCD.

\subsection{The case of $m_u > m_d$}
\label{sec:mumoremd}

It is natural to ask whether the cancellation that occurs precisely at $E/N = 8/3$ as predicted by minimal GUTs is an accident of the low--energy theory of QCD, or whether this has a more fundamental explanation. To address this question, let us consider the very same setup as \eqref{eq:apigamma}, but with the freedom of taking the up quark heavier than the down quark, namely $z > 1$ with the notation of Sec.\,\ref{sec:setup}. As noted in\,\cite{Agrawal:2023sbp}, the monodromic axion charge changes from the case of $z<1$, indicating that the lowest energy path taken by the pion in the axion background is no longer the same. As we shall see, $z > 1$ will actually shift the cancellation discussed in the previous section to a different value of $E/N$. 

\medskip
For $z >1$ it is convenient to take $\tilde Q_a = \,\text{diag}\,(0,1)$. Then, the $\pi^0$ profile satisfying $\partial_{\pi^0} V(a,\pi^0)_{\tilde Q_a} = 0$ is given by:
\be
\pi^0= -\,\text{tan}^{-1} \frac{ \text{sin}( 2N a)}{z+\,\text{cos}\,(2Na)}. 
\ee
The non--linear coupling for $z>1$ becomes
\be
g(a)_{z > 1} = -\,\text{tan}^{-1} \frac{ \text{sin}(2 N a)}{z+\,\text{cos}\,(2 N a)} + \left( E - \frac{2}{3} N\right) a,
\ee
and across any of the possible axion domain walls one has:
\be
\Delta[ g(a)]_\text{DW}^{z > 1} = \left( \frac{E}{N} - \frac{2}{3} \right) \pi.
\ee
As we can see, there is nothing inherently special about $E/N = 8/3$ as the cancellation discussed in the previous section would already be spoiled by the inverted mass hierarchy for the up and down quarks. 
A cancellation can nevertheless still occur, now for a different value of $E/N =2/3$, leading to a qualitatively similar phenomenology. 
We then conclude that axion (and ALP) domain walls can in general become transparent to light for a certain value of $E/N$, which is however model dependent.

\subsection{Heavy QCD axion}
\label{sec:heavy}

In this section we consider the case of a heavy QCD axion that receives two contributions to its potential, namely from QCD and from an additional confining sector. The two contributions can be aligned such that the heavy QCD axion still solves the strong CP problem, see e.g.\,\cite{Tye:1981zy,Berezhiani:2000gh,Dimopoulos:2016lvn,Gherghetta:2016fhp,Agrawal:2017ksf,Hook:2019qoh,Gherghetta:2020ofz}.

In the following, we shall assume that the new confining sector is not charged under the SM gauge group, and that the potential in \eqref{eq:Videntity} is modified by including an additional contribution that involves the axion field only,
\be
\label{eq:heavy}
\tilde V(a, \pi^0) = V_\text{H}(a) + V(a,\pi^0)_{\mathbb{I}/2}, \quad V_\text{H}(a) = -\Lambda_\text{H}^4 \,\text{cos}( 2 N a),
\ee
where we have taken $V_\text{H}(a)$ to be a cosine for simplicity.

Assuming $\Lambda_\text{H} \gg \Lambda_\text{QCD}$, the axion will be much heavier than the pion. The domain wall solutions with the potential in Eq.\,\eqref{eq:heavy} will have an axion profile which is approximately given by 
\be
\label{eq:heavyaxionprof}
a_\text{w}(z) = \frac{2}{N} \,\text{tan}^{-1}\left( e^{m_a z} \right).
\ee
The pion field is then obtained by solving
\be
-f_\pi^2 \frac{\text{d}^2}{\text{d} z^2} \pi^0(z) + \frac{\partial}{\partial \pi^0} V(a_\text{w}(z),\pi^0(z))_{\mathbb{I}/2} = 0.
\ee
In the following, we shall take the pion profile to be used in the explicit calculations to be
\be
\label{eq:profpi0H}
\pi^0(z) = - 2 \,\text{tan}^{-1} \left( e^{m_\pi z} \right),
\ee
which was shown in Ref.\,\cite{Blasi:2023sej} to be a good approximation in this setup.

Due to the assumed alignment between QCD and the additional confining sector, the minima of $\tilde V(a,\pi^0)$ are actually unchanged compared to the potential in \eqref{eq:Videntity}. However, due to $V_\text{H}(a)$, the axion and the pion wall profiles are now varying over two different scales, namely $m_a^{-1}$ and $m_\pi^{-1}$, respectively, with $m_a \gg m_\pi$. For this reason, the photon scattering can no longer be described by a simple function $g(\pi)$ or $g(a)$ as done in Sec.\,\ref{sec:EFT}. 

For the equations of motion of a photon scattering off the wall, we then refer to \eqref{eq:gammaDW} with $\beta(z)$ given in \eqref{eq:betaidentity}. For the scattering to be reliably computed within the meson potential in chiral perturbation theory, the photon frequency should be less than $\Lambda_\text{QCD}$. According to the heavy axion picture one has $m_a \gg m_\pi$, so that photons will always see the axion component as thin.
The reflection probability in the leading Born approximation is then given by
\be
\label{eq:Rpmheavyinter}
\mathcal{R}_\pm^\text{H}(\omega \ll m_a) \simeq  \frac{\alpha^2}{4 \pi^2} \bigg|  \left( \frac{E}{N} - \frac{5}{3} \right) \pi + \mathcal{O}(\omega^2/m_a^2) + \int_{-\infty}^{+\infty} \text{d} z^\prime \, \frac{\text{d}}{\text{d}z^\prime}[\pi^0(z)] e^{2 i \omega z^\prime} \bigg|^2.
\ee 
where we have used $\Delta a = \pi/N$ across the wall. This expression can be solved analytically with the profile in \eqref{eq:profpi0H}, yielding
\be
\label{eq:Rexact}
\mathcal{R}_\pm^\text{H}(\omega \ll m_a) \simeq  \frac{\alpha^2}{4} \bigg|  \left( \frac{E}{N} - \frac{5}{3} \right) - \,\text{sech}\,(\pi \omega/m_\pi) + \mathcal{O}(\omega^2/m_a^2) \bigg|^2.
\ee
The second term in \eqref{eq:Rexact} takes into account that photons with frequency $\omega \sim m_\pi$ will be able to probe the actual structure of the pion profile. For $\omega \gtrsim m_\pi$ this contribution will be negligible, whereas for $\omega \ll m_\pi$ the result for the reflection probability approaches the one in \eqref{eq:Rgenericlow}. This is because the photon sees both the pion and axion components of the wall as thin, and the reflection probability is then only controlled by the value of the fields on the two sides of the wall. As these values are the same as in Sec.\,\ref{sec:setup} due to the alignment of the confining sectors, one obtains the same result in this limit:
\be
\label{eq:Rgenericlowheavy}
\mathcal{R}_\pm^\text{H} (\omega \ll m_\pi) \simeq \frac{\alpha^2}{4} \left(  \frac{E}{N} - \frac{8}{3} \right)^2,
\ee 
where again the case of $E/N = 8/3$ needs special treatment, 
\be
\mathcal{R}_\pm^{\text{H,}\,8/3}(\omega \ll m_\pi) \simeq c_\text{H} \, \alpha^2 \, (\omega/m_\pi)^4, \quad c_\text{H} = 6.08..,
\ee
which is analogous to \eqref{eq:R83low}.

At intermediate momenta, $m_\pi \lesssim \omega \ll m_a$, the reflection probability approaches a constant again, this time given by
\be
\label{eq:Rgenericlowheavy}
\mathcal{R}_\pm^\text{H} (m_\pi \lesssim \omega \ll m_a) \simeq \frac{\alpha^2}{4} \left(  \frac{E}{N} - \frac{5}{3} \right)^2.
\ee 
The special value of $E/N = 5/3$ leads to a qualitatively different behavior in this regime as the contribution from the heavy axion exactly cancels out, and one is left with reflection off the pion profile only, which is exponentially suppressed.

\begin{figure}
\centering
\hspace{-0.4cm}
 \includegraphics[scale=0.26]{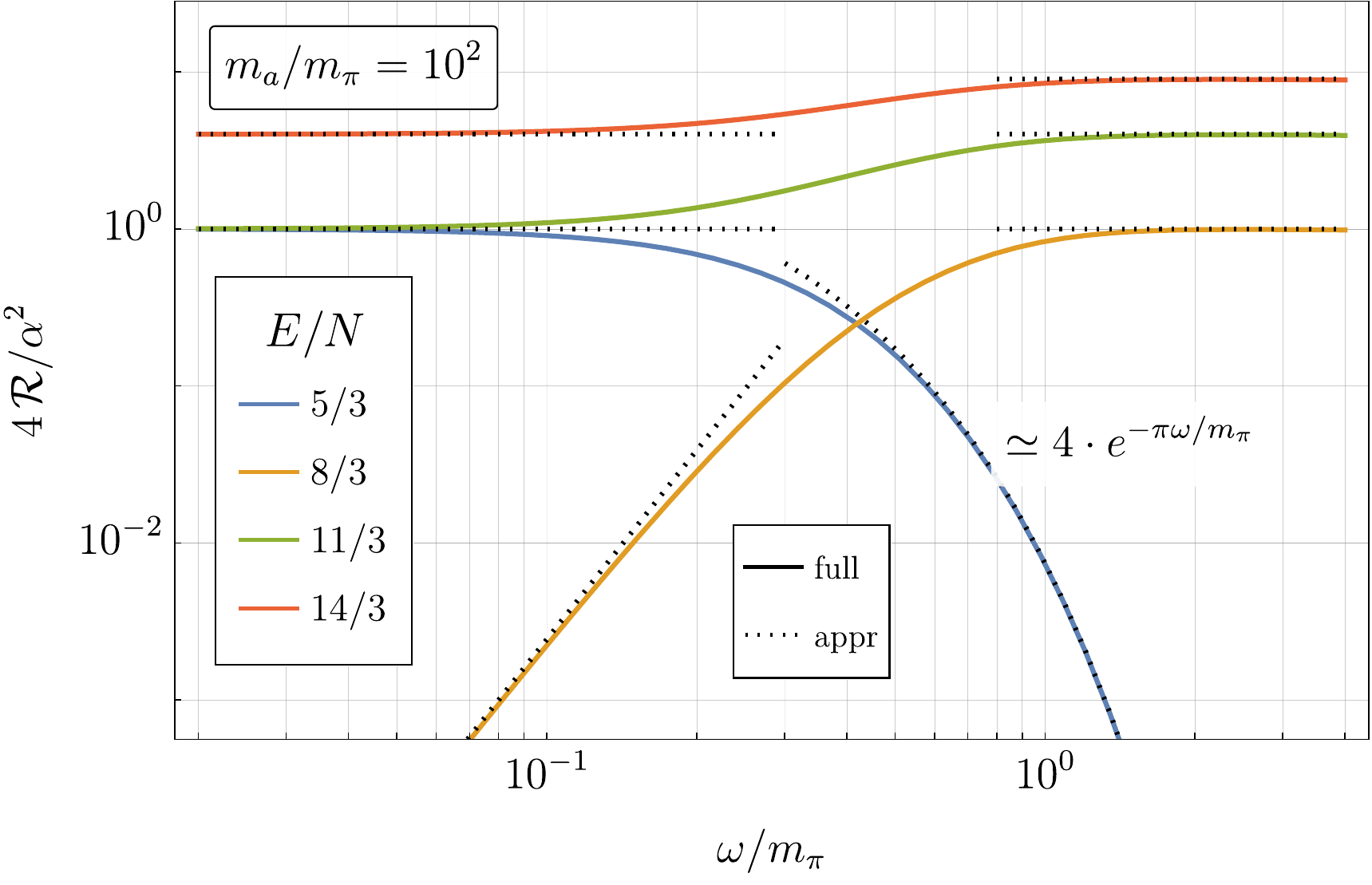}
  \includegraphics[scale=0.26]{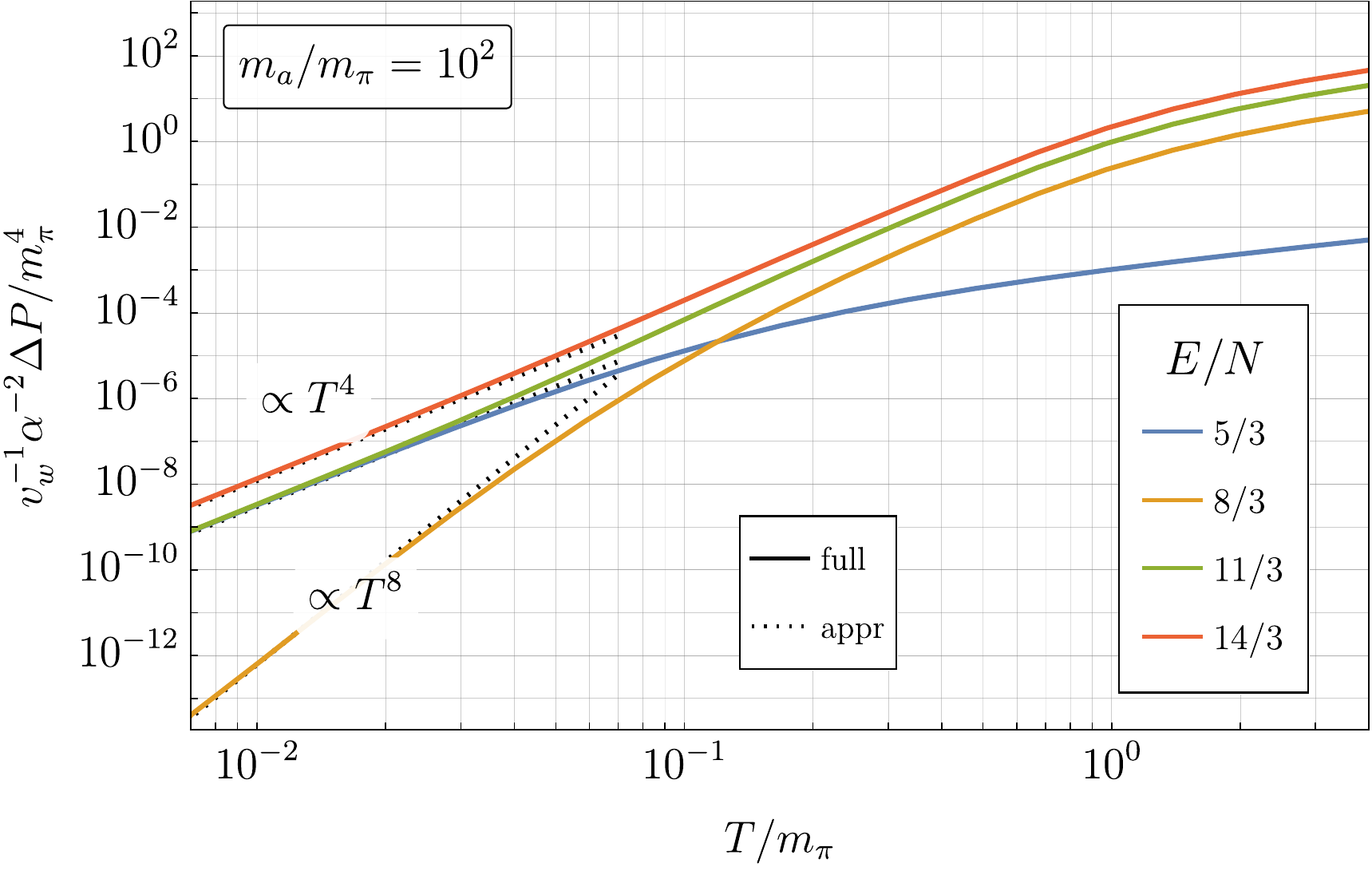}
 \caption{\textbf{Left:} Reflection probability for a photon scattering off the axion--pion domain wall for the case of a heavy QCD axion for different values of $E/N$ as a function of frequency. \textbf{Right:} Thermal pressure on the axion/pion domain wall from a bath of photons for different values of $E/N$. For this calculation we have taken $v_w = 0.1$.}
\label{fig:heavy} 
\end{figure}

\medskip
The reflection probability for different values of $E/N$ is shown in the left panel of Fig.\,\ref{fig:heavy}. As we can see, $E/N = 8/3$ would provide a cancellation at low energies even for the heavy QCD axion case, while $E/N = 5/3$ shows a special behavior in the intermediate kinematic regime as explained above. For other values of $E/N$ the probability is $\mathcal{R} \sim \alpha^2$ and undergoes only a minor shift in the whole kinematic regime of interest ($\omega \ll m_a$). The thermal pressure on the wall is shown in the right panel of Fig.\,\ref{fig:heavy} for the same values of $E/N$. At low temperatures the qualitative behavior is the same as for the standard QCD axion case shown in Fig.\,\ref{fig:fig} with $E/N = 8/3$ leading to $\Delta P \propto T^8$, while for generic values one has $\Delta P \propto T^4$. At intermediate temperatures, $T \gtrsim m_\pi$, the case of $E/N = 5/3$ provides a lower pressure due to the suppressed reflection probability.

\smallskip
At scales much higher than $\Lambda_\text{QCD}$, the domain wall is given only by its heavy axion component. According to the minimal model under consideration, the axion will couple linearly to the photons with a strength given by the electromagnetic anomaly $E$. In this case one simply has:
\be
\mathcal{R}_\pm^\text{H}(\omega) \simeq \frac{E^2 \alpha^2}{4 N^2} \,\text{sech}^2(\pi \omega/m_a),
\ee
where we have used the axion profile as given in \eqref{eq:heavyaxionprof} as well as the leading--order Born approximation.
For small velocities, this leads to a pressure going like $\Delta P \sim v_w T^4$ and $\Delta P \sim v_w m_a^3 T$ for $T\ll m_a$ and $T \gg m_a$, respectively. This in fact corresponds to the case of a linear axion coupling shown by the dashed lines in the right panel of Fig.\,\ref{fig:fig}, which is now accurate for the heavy QCD axion in the kinematic regime under consideration as there is no mixing with meson states.
We however stress again that collective effects from the thermal plasma can modify this picture in the regime $T \gg m_a$\,\cite{Hassan:2024bvb}.

\medskip

Let us conclude this section by noticing that, as long as the heavy axion--pion wall is well established at temperatures below $\Lambda_\text{QCD}$, the birefringent properties will be the same as given in Sec.\,\ref{sec:biref}, namely:
\be
\label{eq:dPhiDWheavy}
\Delta \Phi_\text{DW}^\text{H} =  \pm \frac{\alpha}{2} \left( \frac{E}{N} - \frac{8}{3} \right).
\ee

\section{Conclusion}

The low--energy mixing of the QCD axion with mesons 
can qualitatively modify the electrodynamics of axion defects.
This can be conveniently taken into account by considering the non--linear axion coupling to photons, $g(a) F \tilde F$, which plays an important role for strings and walls where the axion experiences an $\mathcal{O}(1)$ excursion of its fundamental period. 

In this spirit, we have revisited the calculation of photons scattering off axion domain walls first presented in Ref.\,\cite{Huang:1985tt}, clarifying the nature of the precise cancellation for the reflection probability that takes place for the minimal GUT prediction of the PQ electromagnetic and color anomaly, $E/N = 8/3$. For this particular value of $E/N$, the photon friction on the axion wall at low temperatures (where the wall can be considered thin) follows a special power law with the temperature rather than exponential suppression:
\begin{equation*}
\Delta P^{\, 8/3} \sim v_w \, \alpha^2 \left(T/m_a\right)^4 T^4 \quad \left[ T\ll m_a, \,\, v_w \ll 1\right],
\end{equation*}
whereas away from this special cancellation, the generic prediction is:
\begin{equation*}
\Delta P^{\, E/N} \sim v_w \, \alpha^2 \left(\frac{E}{N} - \frac{8}{3}\right)^2 T^4 \quad \left[ T\ll m_a, \,\, v_w \ll 1\right].
\end{equation*}
The analogous expressions for $\Delta P$ in the high--temperature regime can be found in Sec.\,\ref{sec:thermal}.

By comparing the results obtained with the non--linear coupling $g(a)$ against the ones with its linearized version, $\frac{1}{4} g_{a \gamma \gamma} a F \tilde F$, we have shown how the latter does not in general provide an accurate description of the defect dynamics, with the largest discrepancy found for the special value $E/N = 8/3$. These results highlight how the electromagnetic properties of axion particle excitations and axion defects can in principle be very different from each other, as for instance the linear coupling $g_{a \gamma \gamma}$ may be tuned to zero while having $\Delta[ g(a)] = \mathcal{O}(1)$, and vice versa.

A similar observation can be made when considering the birefringent properties of the axion walls, for which we have extended the calculation beyond the WKB approximation of a slowly varying axion field, finding the general result:
\begin{equation*}
\Delta \Phi_\text{DW} = \frac{\alpha}{2\pi} \Delta[g(a)]_\text{DW}.
\end{equation*}

Finally, by considering variations of the standard QCD axion setup, we have noticed that there is nothing inherently special about the value $E/N = 8/3$, as the corresponding cancellation would already be spoiled by the inverted quark mass hierarchy, $m_u > m_d$. Nevertheless, an equivalent cancellation would still take place in this case for a shifted value of $E/N = 2/3$, showing that axion and more generally ALP domain walls can become transparent to light for certain (model--dependent) values of the anomalies. On the other hand, we have found that the cancellation at $E/N = 8/3$ remains true also for heavy QCD axion models, where the axion receives an additional contribution to its mass from a dark confining sector aligned with QCD.

\acknowledgments
We thank Prateek Agrawal, Xucheng Gan, Yu Hamada, Hyungjin Kim, Arthur Platschorre, and Sam Witte for feedback on the draft and useful discussions. Special thanks to Alberto Mariotti for the help throughout the completion of this project.
SB is supported by the Deutsche Forschungsgemeinschaft under Germany’s Excellence Strategy - EXC 2121 Quantum Universe - 390833306.

\bibliographystyle{JHEP}
{\footnotesize
\bibliography{biblio}}

\end{document}